\documentclass[aps,pra,twocolumn,nopacs,superscriptaddress,groupedaddress]{revtex4}

\usepackage{amsmath}
\usepackage[dvips]{graphicx}
\usepackage{epsfig}
\usepackage{tabularx}
\usepackage{color}
\usepackage[pdfpagemode=UseNone,colorlinks=true,linkcolor=blue,citecolor=blue]{hyperref}
 
\usepackage{soul}
\usepackage{gensymb}
\usepackage{upgreek}

\usepackage{amsmath}

\begin{document}

\title{Optical centrifuge for nanoparticles}

\author{Peiyao Xiong}

\author{Kit Ka Kelvin Ho}
\author{J.M.H. Gosling}%
\author{M. Rademacher}
\author{P. F. Barker}
\email{p.barker@ucl.ac.uk}
\affiliation{Department of Physics \& Astronomy, University College London, London WC1E 6BT, UK}

\date{\today}

\begin{abstract}
We study the creation of an optical centrifuge for the controlled rotation of levitated nanorotors within an optical tweezer. The optical centrifuge is created by rapidly rotating the linear polarization of the tightly focused optical field used to form an optical trap. We show that nanorotors, formed by anisotropic nanoparticles levitated within the trap, can be accelerated to well-defined rotational rates in excess of 100 MHz over durations of hundreds of microseconds. The initial conditions required for stable acceleration, based on optical trap properties and the anisotropic susceptibility of the nanorotor are established, and confirmed by numerical simulations. We also present initial experiments that have developed tools for the rapid angular acceleration of the polarization vector of the linearly polarized beam that is required to create the centrifuge. We show that over the acceleration durations in the 100 $\upmu$s range, high rotational speeds could be achieved in modest vacuum.     
\end{abstract}


\maketitle


\section{\label{sec:level1}Introduction}

The field of levitated optomechanics has seen significant developments over the last fifteen years, and of primary importance has been the demonstration of the cooling of translational trapped motion to the quantum ground state \cite{delic_cooling_2020,piotrowski2023simultaneous,magrini2021realtime,tebbenjohanns2021quantum,ranfagni2022twodimensional,kamba2023Revealing}.  Advances in the control, manipulation and detection of the trapped particles promises the creation of non-classical states of rotational motion, which can be used to explore the quantum-to-classical transition at this large mass scales \cite{oriol_quantum_fluctuations_to_macroscopic_dimensions,doi:10.1126/science.abg3027_quantum_classical}. Cold levitated quantum systems also offer the intriguing possibility to test the role of gravity in quantum mechanics \cite{bose_spin_2017,RevModPhys.97.015003_quantumgravity}. 

While much of the early development in levitated optomechanics has focused on translational degrees of freedom, the control and cooling of rotational or constrained rotational (torsional) motion has also seen significant progress \cite{Hoang2016,kamba_nanoscale_2023,rashid_precession_2018,staanum_rotational_2010,pontin_simultaneous_2023,Ahn2020Ultrasensitive,bang_five-dimensional_2020,PhysRevResearch.6.033009Cooling2}, and recently demonstrated ground state cooling of the torsional motion \cite{dania2024hpqoart}. A range of protocols have been proposed for creating nonclassical rotational motion. This includes the observation of the revivals of rotational wavepackets~\cite{stickler_probing_2018, stickler_quantum_2021} and the quantum tennis racket effect~\cite{ma_quantum_2020}. 

The optical centrifuge is an established tool in molecular physics that is used to induce controlled rotation of molecules to high rotational states \cite{karczmarek_optical_1999}. This mechanism has allowed the creation of narrow rotational wave packets that have been used to study molecular structure, collision processes \cite{michael_state-resolved_2021}, optical alignment \cite{korobenko_adiabatic_2016} and chirality \cite{milner_controlled_2019}. An optical centrifuge transfers angular momentum from an optical field to molecules and is accomplished by applying a constant angular acceleration to the linear polarization vector within a short pulsed optical field \cite{macphail-bartley_laser_2020}.

In this paper, we study the creation of an optical centrifuge for the controlled rotation of nanorotors, levitated within an optical tweezer. This scheme is suitable for experiments that aim to probe the quantum nature of rotational motion of these mesoscopic rotors. We present a description of this process and discuss optimized schemes for acceleration to well-defined rotational frequencies. We also present initial experiments that have developed tools for the rapid angular acceleration of the polarization vector of the linearly polarized tweezer beam that are required to create the optical centrifuge.

\section{Rotational control}

The process of rotating a nanorotor or a molecule can be understood as a scattering process \cite{minck_stimulated_1966}. Rotational Raman scattering is a form of inelastic photon scattering in which a molecule can gain or lose rotational quanta~\cite{noauthor_microwave_nodate,noauthor_rotational-raman_nodate}. A series of these processes can be used to create an optical centrifuge, which exploits Raman transitions to raise a molecule's rotational angular momentum to very high levels and even to dissociation \cite{korobenko_control_2018}. As each photon only has one unit of internal angular momentum, each transition only increases the angular momentum of the molecule $J$ by 2 units. Thus, the rotational ladder has to be climbed successively in a finite time with a controlled frequency chirp to reach the desired rotational level. 

The transfer of angular momentum is a scattering process that is achieved using polarized light. For example, when circularly polarized light is absorbed or scattered by a particle, each photon imparts one unit of momentum, $\hbar$, onto an object \cite{beth_mechanical_1936}. This gives a torque that rotates a particle in the plane perpendicular to the direction of light propagation \cite{Friese1998,PhysRevLett.117.123604_lightprapagation,PhysRevLett.121.033602_GHzrotate,Arita2013_Laser_induced}. However, this approach does not offer precise control over the final angular momentum imparted, as the terminal angular rotation rate is usually determined by dissipation from gas. An alternative method is to first align the nanorotor in a linearly polarized field and then slowly rotate the polarization up to a well-defined angular velocity so that the particle's alignment will adiabatically follow it. This is the concept of an optical centrifuge \cite{opticalcentrifuge}. 

\begin{figure}
    \centering
    \includegraphics[width=\columnwidth]{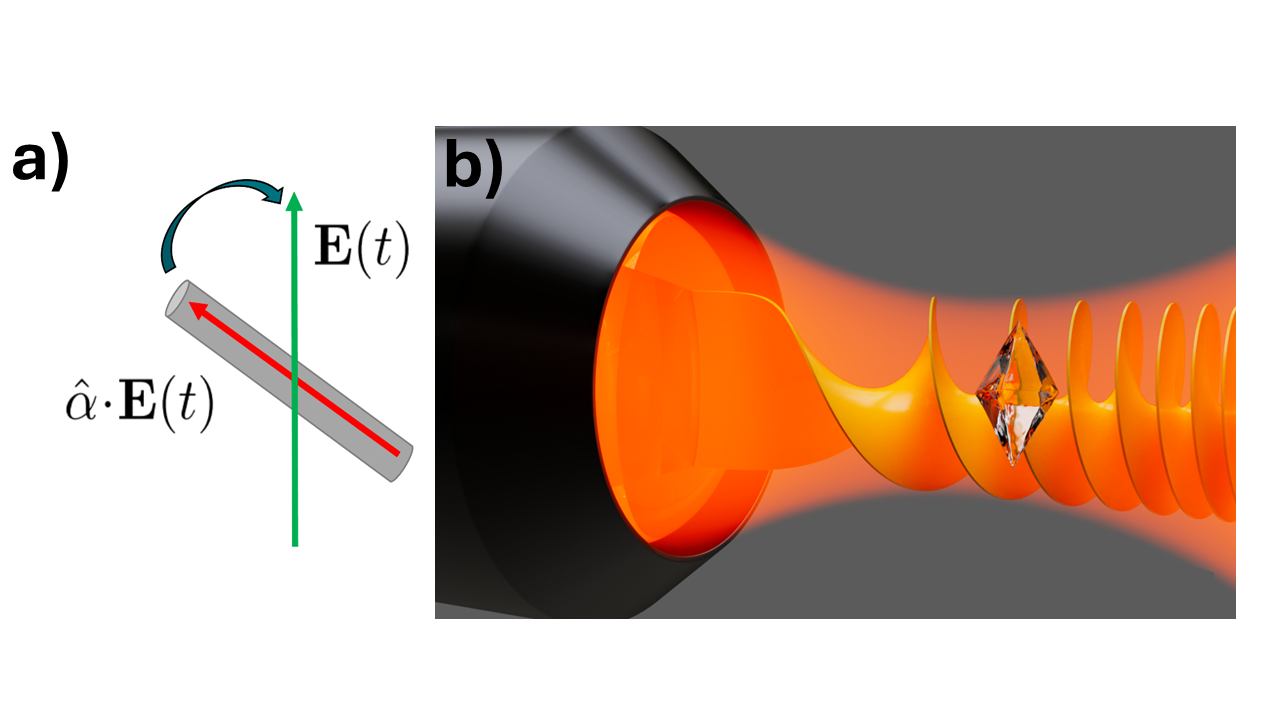}
    \caption{a) Diagram showing the torque on the induced dipole moment (red arrow) of a nanorotor (grey cylinder) with an induced dipole moment due to an electric field shown as the green arrow. b) Illustration of optical centrifuge formed with an optical tweezer by the linear polarization of the tweezers optical field. The nanorotor represented by the bipyramidal structure  is trapped in the linear polarized tweezer beam whose polarization accelerates as a function of time as shown by the yellow spiral. The acceleration of the polarization is slow enough so that a trapped nanorotor adiabatically aligns with the polarization axes. Upon reaching the desired rotation frequency, the optical field is switched off, leaving the rotor spinning at a well-defined angular frequency.}
    \label{fig:optical_centrifuge2}
\end{figure}

This alignment process occurs when the polarization vector of the illuminating optical field induces a dipole moment given by $\mu=\hat{\alpha}\cdot \mathbf{E}(t)$,  where $\hat{\alpha}$ is the polarizability tensor of the rotor. In turn, the induced dipole moment interacts with the electric field, and an optical potential of the nanorotor is formed and given by \cite{stapelfeldt_colloquium_2003}
\begin{equation}
  U = -\frac{1}{2} \langle\mathbf{E}(t) \cdot \hat{\alpha} \cdot \mathbf{E}(t)\rangle.
  \label{eq:induceddipolemoment}
\end{equation}
The minimum of this energy occurs when the largest polarizability of the rotor aligns with the electric field. For an axisymmetric rotor, the long axes will align with the direction of the electric field. When the rotor is not aligned, there would be a torque which pushes it towards alignment creating an oscillating torsional/librational or pendular motion, see Fig. \ref{fig:optical_centrifuge2}(a). This is observed in many optical levitation experiments \cite{hoang_torsional_2018,kamba_nanoscale_2023,pontin_simultaneous_2023}.

Fig. \ref{fig:optical_centrifuge2}(b) is a schematic diagram of an optical centrifuge for a nanorotor. The nanorotor is trapped in the linear polarized light field of an optical tweezer formed by a tightly focused laser beam. In order to rotate up to a well-defined frequency, the rotor must adiabatically align with the evolving light-field polarization. To enable this, the field must first rotate slowly enough so that it drags the rotor along the polarization axes in a slow and controlled manner. The rotation rate is increased, dragging the rotor to higher angular velocities. Once the desired angular/rotational frequency is reached, the light field can be rapidly turned off and its motion can evolve in the absence of light scattering~\cite{korobenko_control_2018}.



\section{\label{sec:level2} Equations of motion for a levitated nanorotor}

\begin{figure}
    \centering
    \includegraphics[width=0.418\textwidth]{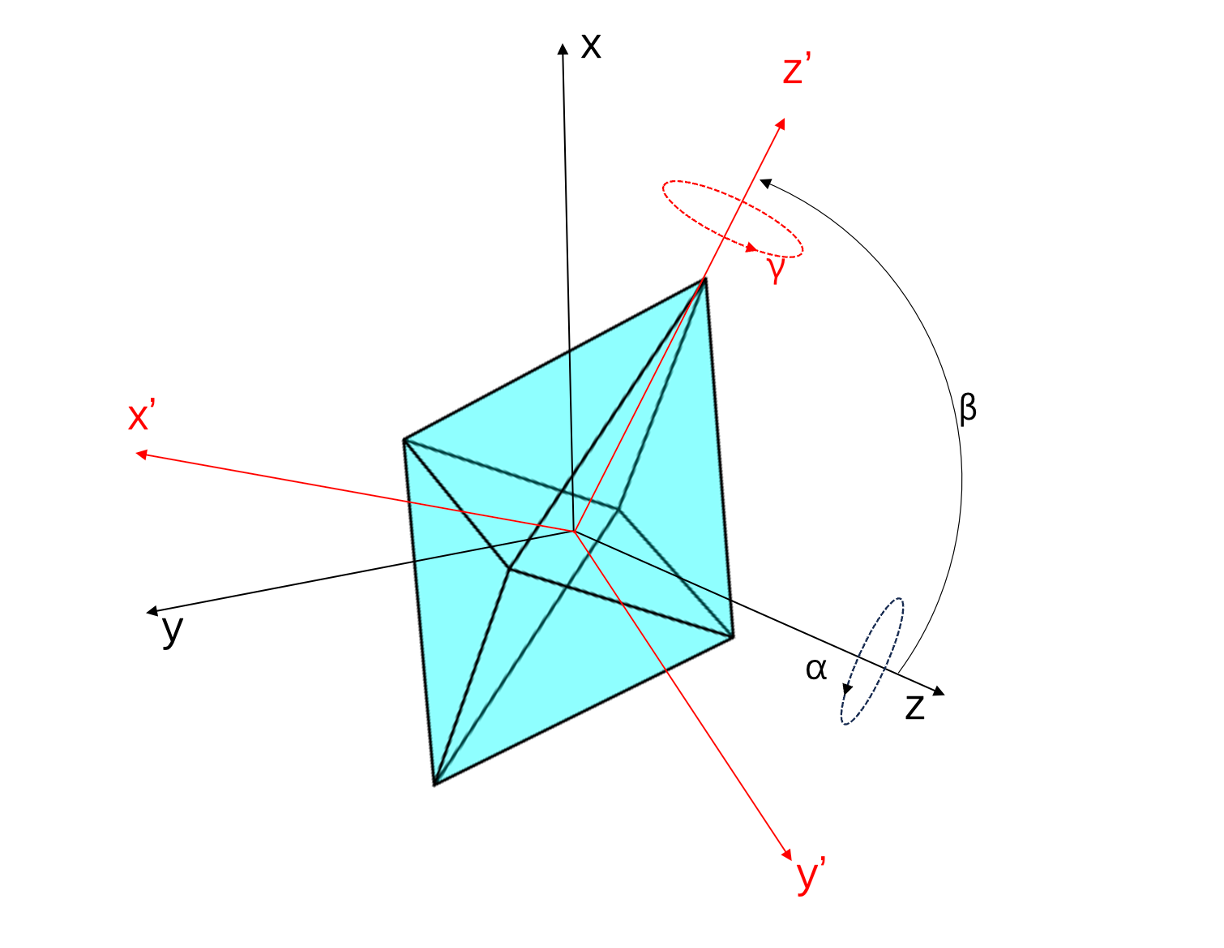}
    \caption{The schematic representation of a nanorotor formed by a bi-pyramidal structure trapped by a laser beam traveling in $z$ direction. The $x,y,z$ are laboratory axes and $x',y',z'$ are body-frame axes. The angles $\alpha,\beta,\gamma$ are defined in the Euler convention, where $\alpha$ is the rotational angle about the laboratory z axis, $\beta$ represents the angle between the laboratory $z$ axis and the body-frame $z'$ axis, the rotational angle about $z'$ axis is denoted by $\gamma$. } 
    \label{fig:nanoparticle}
\end{figure}

To study an optical centrifuge for nanorotors, we consider a nanoparticle trapped by a polarized Gaussian laser beam. As shown in Fig. \ref{fig:nanoparticle}, the rotation is considered in the Euler convention~\cite{rashid_precession_2018}. The nanoparticle has the electric susceptibility tensor $\chi= \text{diag}(\chi_1, \chi_2, \chi_3)$ and the moment of inertia $J = (J1, J2, J3)$ in the body-frame axes $x',y',z'$. 


The motion of nanoparticles under the influence of the tweezer fields can be modeled using the Hamiltonian below \cite{Hgrad}

\begin{equation}
    H=H_\mathrm{grad}+H_\mathrm{free}
    \label{Htotal}
\end{equation}
where $H_\mathrm{grad}$ and $H_\mathrm{free}$ represent the energy due to the gradient in the optical potential and the free rotor energy respectively. As we consider nanorotors smaller than the wavelength of light, we do not consider the effect of scattering forces.

The term due to the gradient optical force can be determined by

\begin{equation}
    {H}_{\mathrm{grad}}=-\frac{V P}{c \sigma_L}|{u}|^2\left(\boldsymbol{\epsilon}_d^*\right)^{\top} {\hat{\chi}} \boldsymbol{\epsilon}_d
    \label{Hgrad}
\end{equation}
where $V$ is the volume of the nanoparticle, $P$ is the laser power, $c$ is the speed of light, $\hat{\chi} = R \chi R^\top$is the electric susceptibility tensor in the lab frame, $\epsilon_d$ is the polarization vector, $\sigma$ is the transverse cross-section of the beam: $\sigma_L=\pi w_0^2$, ${u}$ represents the spatial variation of a Gaussian optical field which has the form in the focal region:

\begin{equation}
    {u}=\frac{w_0}{w({z})}e^{-\frac{1}{\omega(z)^2}\left(\frac{x^2}{\omega_x^2}+\frac{y^2}{\omega_y^2}\right)}e^{i k {z}}
\end{equation}
where $w_x$ and $w_y$ are the beam waists in $x$ and $y$ direction, $k$ is the angular wave number, $w(z)=w_0 \sqrt{1+(z/z_R)^2}$, $w_0$ is the beam waist and $z_R = \pi \omega_0^2/\lambda$ is Rayleigh range of the beam. The Hamiltonian that includes all Euler angles is given by \cite{Hgrad}


\begin{equation}
\begin{aligned}
H_{\mathrm{grad}}=
& -\frac{V P}{c \sigma_L}\lvert u\rvert^2 \\
&
\Bigl(b_x^2\Bigl[\chi_1(\cos(\alpha)\cos(\beta)\cos(\gamma) - \sin(\alpha)\sin(\gamma))^2 \\
&\qquad
+ \chi_2(\cos(\alpha)\cos(\beta)\sin(\gamma) + \sin(\alpha)\cos(\gamma))^2 \\
&\qquad
+ \chi_3 \cos^2(\alpha)\,\sin^2(\beta)\Bigr] \\
&
+ b_y^2 \Bigl[\chi_1(\sin(\alpha)\cos(\beta)\cos(\gamma) + \cos(\alpha)\sin(\gamma))^2 \\
&\qquad
+ \chi_2(\cos(\alpha)\cos(\gamma) - \sin(\alpha)\cos(\beta)\sin(\gamma))^2 \\
&\qquad
+ \chi_3 \sin^2(\alpha)\,\sin^2(\beta)\Bigr]
\Bigr).
\label{Hgradex}
\end{aligned}
\end{equation}

The free Hamiltonian is given by \cite{rashid_precession_2018}
\begin{equation}
    H_{\text {free }}=\frac{P_x{ }^2+P_y{ }^2+P_z{ }^2}{2 M}+\left(\frac{\Pi_1(\beta, \gamma)}{2 J_1}+\frac{\Pi_2(\beta, \gamma)}{2 J_2}+\frac{\pi_\gamma{ }^2}{2 J_3}\right)
    \label{Hfree}
\end{equation}

where $M$ is the total mass of the system, and $P_{x,y,z}$ and $\pi_{\alpha,\beta,\gamma}$ are the momentum and angular momentum respectively. $\Pi_1$ and $\Pi_2$ are the rotational angular kinetic terms given by

\begin{equation}
\begin{aligned}
    &\Pi_1(\beta, \gamma)=\\
&\csc ^2(\beta)\left[\cos (\gamma)\left(\pi_\alpha-\pi_\gamma \cos (\beta)\right)-\pi_\beta \sin (\beta) \sin (\gamma)\right]^2 \\
\end{aligned}
\end{equation}

\begin{equation}
\begin{aligned}
    &\Pi_2(\beta, \gamma)=\\
    &\csc ^2(\beta)\left[\sin (\gamma)\left(\pi_\alpha-\pi_\gamma \cos (\beta)\right)+\pi_\beta \sin (\beta) \cos (\gamma)\right]^2
\end{aligned}
\end{equation}

By determining the Hamiltonian of the system, we obtain the equation of motion for each translational degree of freedom  $r = (x,y,z)$ as

\begin{equation}
    \frac{\partial r}{\partial t}=\frac{\partial H}{\partial P_r}
    \label{drdt}
\end{equation}
The equation of motion for the angular degrees of freedom, where $\phi=(\alpha, \beta, \gamma)$ can be written as

\begin{equation}
    \frac{\partial \phi}{\partial t}=\frac{\partial H}{\partial P_{\phi}}
    \label{dphidt}
\end{equation}


\section{\label{sec:level3}Dynamics in the accelerated frame}

We consider a centrifuge for a symmetric nanorotor created by rotating the linear polarization of a focused Gaussian beam. We assume that the nanorotor is trapped by a tightly focused laser beam. The rate of rotation of a linearly polarized field that induces torque on the nanorotor is given by $\beta_c = \frac{d\omega}{dt}$ where $\omega$ is the angular velocity of the polarization vector. For a linear increase in angular velocity of the polarization, the angular frequency of rotation is given by $\omega = \omega_0 + \beta_c t $ where $\omega_0$ is the initial angular frequency of the rotor, which we arbitrarily set it to zero. The torque ($\tau$) on the particle can be calculated from
\begin{equation}
    \tau=\frac{\partial H}{\partial \phi}
    \label{tauHphi}
\end{equation}
For a symmetric rotor ($\chi_1 = \chi_2$, $J_1 = J_2$) with volume $V$, the angular acceleration on each degree of freedom, $\phi=(\alpha, \beta, \gamma)$, is given by

\begin{equation}
    a_{\alpha}=-\frac{I_e V z_R^2 (\chi_1-\chi_3) \sin \left[\beta_c t^2-2 \alpha\right] \sin [\beta]^2}{2 c\left(z^2+z_R^2\right) \epsilon_0J_1}
\end{equation}
\begin{equation}
    a_{\beta}=\frac{I_e V z_R^2 (\chi_1-\chi_3) \cos \left[\frac{\beta_c t^2}{2}-\alpha\right]^2 \sin [2 \beta]}{2 c \left(z^2+z_R^2\right) \epsilon_0J_2}
\end{equation}
\begin{equation}
    a_{\gamma}=0
\end{equation}
where $I_e$ is the intensity of an elliptical Gaussian beam is given by      
\begin{equation}
    I_e=I  e^{-\frac{2}{\omega(z)^2}(\frac{x^2}{\omega_x^2}+\frac{y^2}{\omega_y^2})}
\end{equation}
where $I$ is the peak intensity at the centre of the beam. We consider a stationary particle that is cooled such that it is initially at the centre of the beam ($x, y, z, p_x, p_y, p_z \approx 0$). The simplest case is for when the long axis of the rotor is aligned along the polarization direction such that $\beta=\pi/2$. The angular acceleration in $\alpha$ becomes

\begin{equation}
    a_{\alpha}=a_0 \sin \left[\beta_c t^2-2 \alpha\right],
\end{equation}
where the peak angular acceleration along the $\alpha$ axis is
\begin{equation}
    a_0=\frac{I V (\chi_1-\chi_3)}{2 c \epsilon_0J_1}.
    \label{tau0}
\end{equation}

For this simple case there is no acceleration along $\beta$ and $\gamma$ such that $ a_{\beta}=a_{\gamma}=0.$
The phase of the rotor in the accelerated reference frame of the polarization vector is $\theta=2 \alpha-\beta_c t^2$, where the phase velocity $\eta$ in this frame is
\begin{equation}
    \eta=\frac{d \theta}{d T}
    \label{eta}
\end{equation}
and $T=\sqrt{\beta_c} t$. The angular acceleration is
\begin{equation}
    \frac{d \eta}{d T}=-\frac{2 a_0}{\beta_c} \sin \theta-2.
    \label{deat}
\end{equation}
This is analogous to the equation of motion for the acceleration of a particle in a chirped optical lattice \cite{barker_optical_2001}. 



\begin{figure}
    \centering
    \begin{minipage}[b]{\columnwidth}
        \includegraphics[width=\columnwidth]{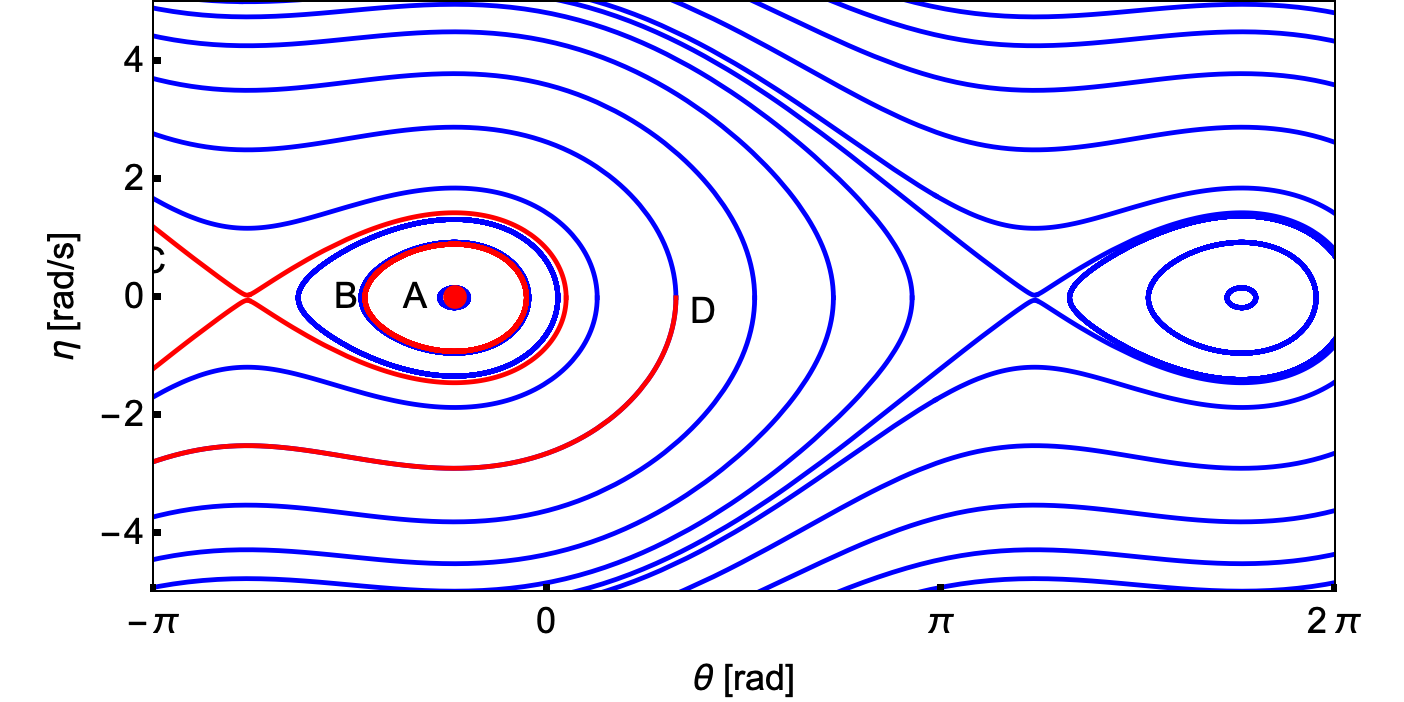}
        \footnotesize \hspace{4 mm} (a)
        \label{fig:ps1}
    \end{minipage}
    \hfill
    \begin{minipage}[b]{\columnwidth}
        \includegraphics[width=\columnwidth]{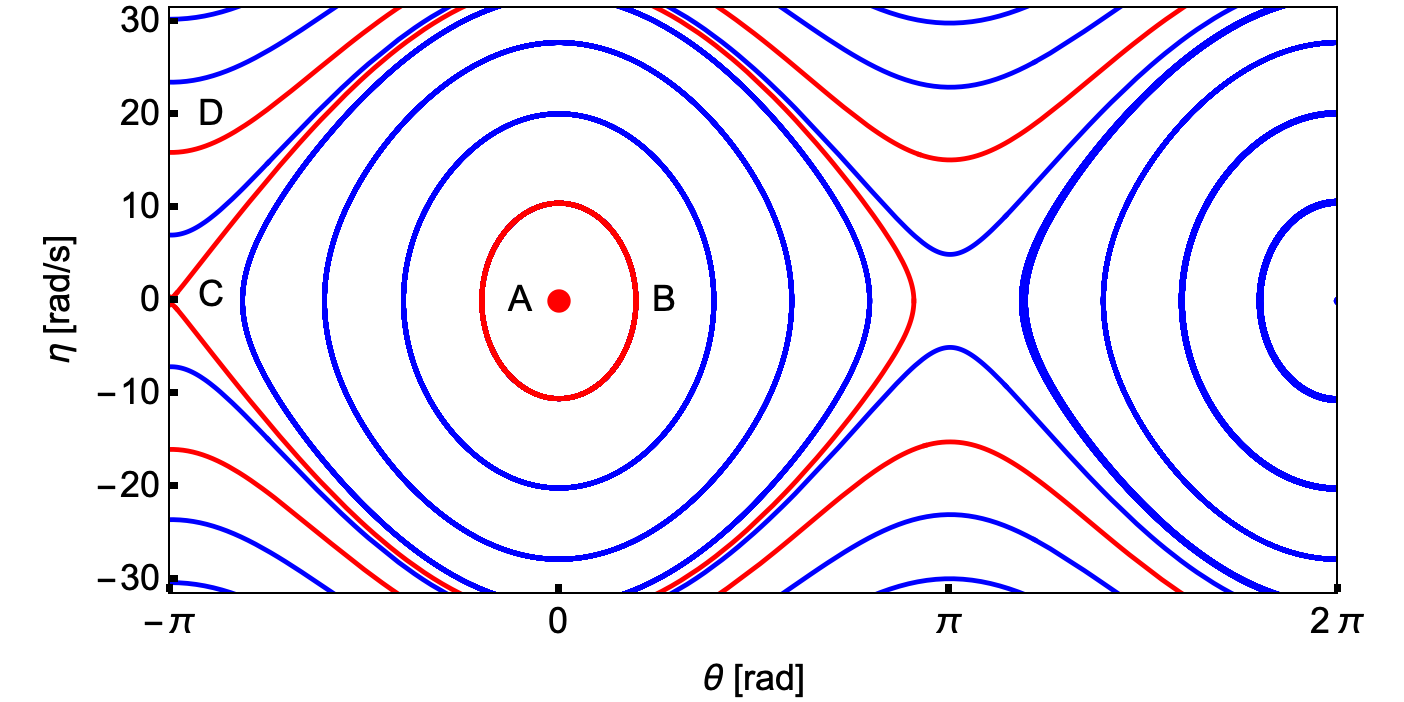}
        \footnotesize \hspace{4 mm} (b)
        \label{fig:ps2}
    \end{minipage}
    \caption{The phase-space ($\theta,\eta$) diagram for trapped particles with different $\psi$ values in accelerated reference frame. (a) Plot for the stability parameter $\psi=\frac{\beta_c}{a_0}=0.68$ where A, B, C and D represent the trajectories of particles with initial phases of $[\theta,\eta]=[-0.75, 0]$, $[\theta,\eta]=[-0.1, 0]$, $[\theta,\eta]=[0.07, 0]$ and $[\theta,\eta]=[1, 0]$ respectively. (b) Plots for the same initial phases as $\psi<0.01$ .}
    \label{fig:phase-space}
\end{figure}

To understand the angular motion in the accelerated frame of a nanorotor,  we plot the trajectories in the $[\theta,\eta]$ phase space as shown in Fig. \ref{fig:phase-space}. The closed tear drop shaped orbits such as trajectory B are when the rotor is trapped by the accelerating rotating potential while trajectory C and D show untrapped motion and do not follow the accelerating angular potential. Point A is the critical point in which the particle exactly follows the accelerating potential and there is no librational motion in that reference frame. The critical angle can be determined by setting equation \ref{eta} and \ref{deat} to be zero such that
\begin{equation}
    \sin \theta_c=-\psi
    \label{sintheta}
\end{equation}
where $\psi= \beta_c / a_0$ is called the stability parameter. The critical points correspond to $[\theta,\eta]=[\theta_c+2n \pi, 0]$, where $n$ is an integer. Moreover, according to equation \ref{sintheta}, in order for the critical angle to be real, the stability parameter must be
\begin{equation}
    |\psi| < 1
    \label{psilim}
\end{equation}
for the rotor to follow the accelerating potential.  

As an example, we consider the requirements to drive a bipyramidal nanorotor, as shown in Fig. \ref{fig:nanoparticle}, up to a rotational frequency of 100 MHz. We use a typical peak laser tweezer intensity of a Gaussian beam for trapping nanoparticles at $2.14 \times 10^{11}$ W/m$^2$, assuming the wavelength of laser is $\lambda = 1064$ nm, a beam power of 225 mW and a focused beam waist of $\omega_0 = 818.46$ nm. From equation \ref{tau0}, this gives a peak angular acceleration of $a_0 = 8.85 \times 10^{12}$ rad/s$^2$, and from equation \ref{psilim}, the value of chirp ($\beta_c$) should be less than this value. We consider a chirp of $6 \times 10^{12}$ rad/s$^2$ where $\psi = 0.68$ and the critical angle is $\theta_c = -0.74$ rad, so that we can accelerate the rotor to 100 MHz over a duration of 0.11 ms. This corresponds to a nanorotor orientation that lags maximum alignment where the initial $\alpha = \theta_c/2$.  

\begin{figure}
    \centering
    \includegraphics[width=\columnwidth]{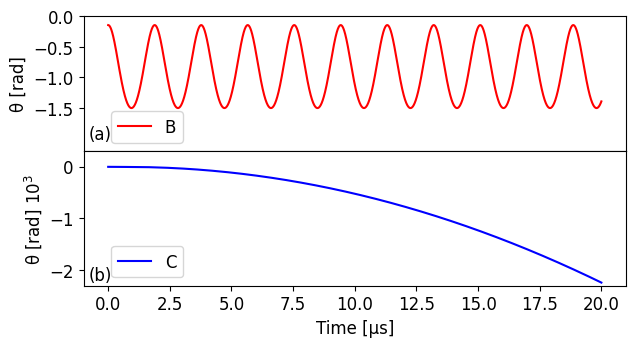}
    \caption{The phase-time diagram of the trapped particles in $\alpha$ direction. (a) Trapped particle, where the red curve corresponds to trajectory B in Fig. \ref{fig:phase-space}(a). For a trapped particle, the phase of it can be constant or oscillate between the equilibrium value. (b) The situation that the particle is lost during the driving process. The blue curve corresponds to trajectory C in Fig. \ref{fig:phase-space}(a). In this situation, the phase value will keep decreasing and the particle can not be accelerated.}
    \label{fig:theta}
\end{figure}

\begin{figure}
    \centering
    \includegraphics[width=\columnwidth]{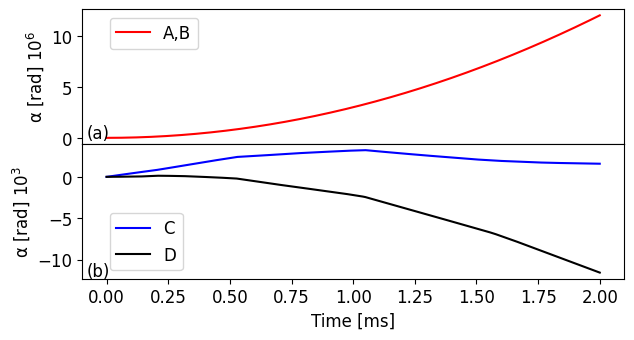}
    \caption{Rotational angle $\alpha$ versus time. (a) Trapped particle, where the red curve represents the particle is initially close to the critical point, which corresponds to trajectory A and B in Fig. \ref{fig:phase-space}(a). (b) The situation that the particle is lose during the driving process. The blue and black curve correspond to trajectory C and D in Fig. \ref{fig:phase-space}a.}
    \label{fig:alpha}
\end{figure}

Fig. \ref{fig:theta} shows the evolution of $\theta$ of the nanorotor in the accelerating frame over time of 20 $\upmu$s. The motion of A is not shown as it is a straight line at 0.74 rad. However, for trajectory B, the particle oscillates. This is librational motion in this reference frame. As trajectory C is untrapped, its angle rapidly decreases with time, as does D but is not shown here. Fig. \ref{fig:alpha}(a) shows the evolution of angle $\alpha$ in the lab frame driven by the optical centrifuge over 2 ms, when its initial phase is close to the critical point (trajectory A and B). The particle is accelerated to a high rotational frequency of nearly 2 GHz. For untrapped trajectory C, $\alpha$ initially increases for about 1 ms but no longer accelerates after this. When the initial position is far from the critical point (trajectory D), the value of $\alpha$ just decreases as it experiences angular acceleration in the opposite direction.  

\begin{figure}
    \centering
    \includegraphics[width=\columnwidth]{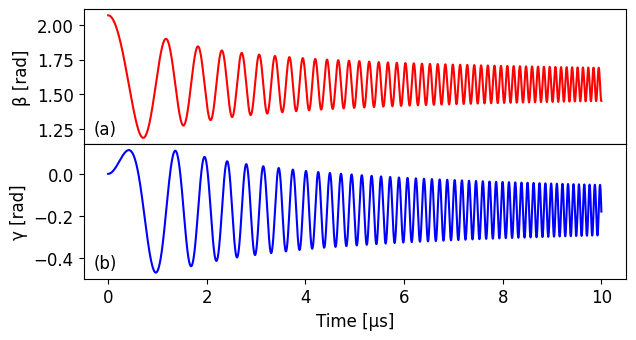}
    \label{fig:subfig6}
    \caption{The rotational angle $\beta$ versus time of a trapped particle with $\beta_0 = \pi/2 + 0.5$ and initially $\alpha$ is at critical point and $\gamma_0 = 0$.  The value of $\beta$ oscillate around the equilibrium point in $\beta$ direction with decreasing amplitude and increasing frequency. }
    \label{fig:betatime}
\end{figure}
We now consider the case where $\beta$ is not at the equilibrium point $\beta = \pi/2$. Fig. \ref{fig:betatime} shows the evolution of $\beta$ over 10 $\mu$s in the optical centrifuge when the nanorotor is 0.5 rad behind the equilibrium point. The value of $\beta$ oscillates around the equilibrium point during acceleration and gets closer to the equilibrium point as time increases. This behavior is independent of the initial value of $\beta$ and $\gamma$ and shows that when a particle is driven in the $\alpha$ direction, its $\beta$ and $\gamma$ angles are couple with each other, as also observed in gyroscope motion \cite{khodorkovsky2015collisional}.





\section{\label{sec:level5}The effect of collisions}

\begin{figure}
    \centering
    \includegraphics[width=\columnwidth]{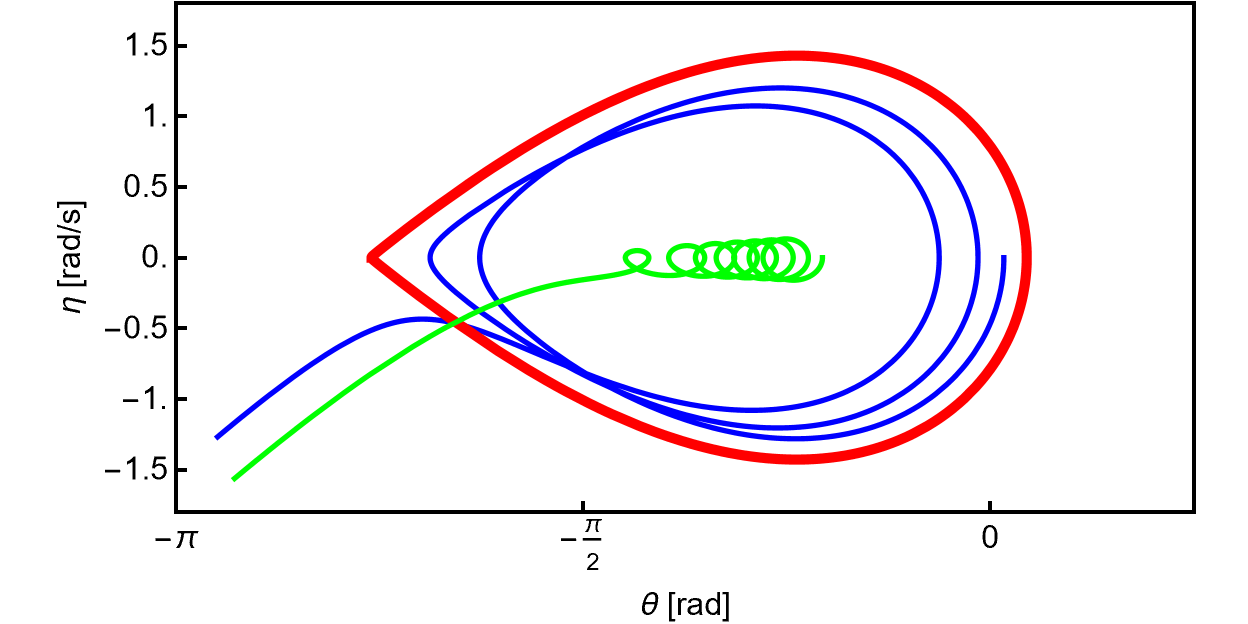}
    \caption{Plot of the dynamics of the nanorotor in the accelerated reference frame in the presence of drag $\gamma$ for different initial phases  $\theta(0)$ = 0.052 (blue) and   $\theta(0)$ =-0.648 (green) and $\eta=0$, for $\psi=0.680$, $\frac{\gamma}{\sqrt{\beta_c}}=0.026$. The thicker red tear drop shaped region represents the boundary between stable motion which follows the acceleration angular potential. Motion within this region is trapped by the optical potential and undergoes angular acceleration. In the presence of damping all particles are eventually lost from the potential. The time that they stay trapped is approximately the same for any initial phase of the nanorotor and given by $\gamma^{-1}$   }
    \label{fig:drag}
\end{figure}

A nanorotor will collide with gas particles and be subjected to the recoil of photons, resulting in a net random torque that leads to a drag torque that increases with time. This will eventually become large enough to push the particle out of the confining potential. In the accelerated reference frame with damping $\gamma_g$, equation \ref{deat} becomes
\begin{equation}
    \frac{d \eta}{d T}=-2(\frac{ a_0}{\beta_c} \sin \theta-1-\frac{\gamma_g} {2\sqrt{\beta_c}}\eta-\frac{\gamma_g} {\sqrt{\beta_c}}T).
    \label{deat2}
\end{equation}
Fig. \ref{fig:drag} is a graph of the nanorotator that experiences drag in the accelerated frame, where  $\frac{\gamma_g}{\sqrt{\beta_c}}=0.0255$ and  $\frac{ a_0}{\beta_c}$=1.47 for two initial phases of $\theta = - 0.0522$ (blue) and $\theta = -0.647763$ (green) and initially stationary in the reference frame. In the presence of drag, the time for the particle to fall out of the optical potential, defined by the red teardrop-shaped region, is largely independent of the initial phase and takes approximately $\gamma^ {-1}$.

This timescale is also confirmed when we model this acceleration in the presence of drag on both position and angular degrees of freedom, as well as the inclusion of stochastic forces and torques, via the Stratonovich processes given by \cite{lemons_introduction_2003}
\begin{equation}
    d P_r = \left(-\frac{\partial H}{\partial r}-\gamma_r P_r\right) \mathrm{dt} + \left({2 k_B T m\gamma_r}\right)^{\frac{1}{2}} \circ \mathrm{d} W_r
\end{equation}

\begin{equation}
    d P_\phi = \left(-\frac{\partial H}{\partial \phi}-\gamma_\phi P_\phi\right) \mathrm{dt} + \left({2 k_B T J_\phi\gamma_\phi}\right)^{\frac{1}{2}} \circ \mathrm{d} W_\phi
\end{equation}
where $W_{r,\phi}$  are the noise terms due to gas collisions only and $\gamma_r$, $\gamma_\phi$ are the translational and rotational damping rates respectively. 

\begin{figure}
    \centering
    \includegraphics[width=\columnwidth]{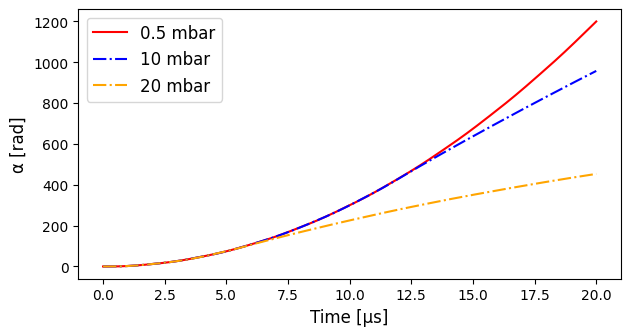}
    \caption{Rotational angle $\alpha$ versus time for different pressures with a chirp rate of  $\beta_c = 6 \times 10 ^{12}$ rad/s$^2$. The initial rotational angle for all three ismulations is  $\alpha$  is -0.365 rad which displaced by 0.005 rad  from the critical point, $\beta$ is $\pi/2 + 0.05$, and $\gamma$ is 0. The red, blue and orange curve represents the trapped particle under 0.5 mbar, 10mbar and 20mbar pressure respectively.}
    \label{fig:differentP}
\end{figure}

The translational and rotational damping coefficients ($\beta_{tr}$ and $\beta_{rot}$) for nanoparticles at low pressures in the Knudsen region can be calculated using the formalism of  by Cavalleri et al \cite{cavalleri_gas_2010}. The translational damping rates are given by

\begin{equation}
    \gamma_{r} = \frac{\beta_{tr} P_g}{v_g m_0}  
\end{equation}
and the rotational damping rates by
\begin{equation}
    \gamma_{\phi} = \frac{\beta_{rot} P_g}{v_g J_\phi}  
\end{equation}
where $P_g$ is the pressure, $v_g = \sqrt{  k_B T /( m_g)}$ is the thermal velocity of the gas and $m_0$ is the mass of the trapped particle. 
Fig. \ref{fig:differentP} shows the evolution of the angle $\alpha$ with different pressures where the motion is dominated by collisions with gas molecules. The nanoparticle's $\alpha$ axis is initially aligned with the polairzation vector so that $\alpha$ = -0.365 and $\beta$ = 1.6208 is shifted by 0.05 rad from its equilibrium point of $\beta_0 = \pi/2$ . We also assume that the nanorotor is initially cooled to 1 K in all degrees of freedom, which can be easily achieved \cite{dania2021opticalCooling1,PhysRevResearch.6.033009Cooling2,pontin_simultaneous_2023,bang_five-dimensional_2020}. This indicates the particles can be accelerated briefly, even at fairly poor vacuum with the time available for acceleration given again by approximately $\gamma ^{-1}$.

For eventual quantum applications, even a single collision should be avoided, and we require that the acceleration duration be significantly less than the average time between collisions for a particle with collisional cross-section $\sigma$.  For a target final rotational  frequency of  $f_t$, and the chirp of the driving polarization, $\beta_c$, the time of acceleration $f_t/\beta_c$ should obey the inequality,

\begin{equation}
    \frac{f_t}{\beta_c} < \frac{2 k_B T}{\pi \sigma v_g P}.
    \label{flim}
\end{equation}

Using this expression we find that for rotational frequency of 100 MHz and a final chirp rate of $\beta_c = 6 \times 10 ^{12}$ rad/s$^2$ we require a pressure below approximately 10$^{-6}$ mbar. This is easily achieved in experiments.




\section{Electro-Optical polarization Control}
Rapid changes in polarization cannot easily be achieved using bulk electro-optical crystals as fast switching of high voltages is required. To perform the required polarization control change at high speeds, it is more practical to use optically integrated  polarization controllers with a birefringent wave guide that is voltage controlled. Light is coupled into the device by single-mode fibers and only requires voltages in the $\pm$30 volt range for complete polarization control. Additionally, as we seek to rotate the particle at MHz frequencies, the device will also need to have a response time at a comparable frequency, and as such its low response time makes it perfectly suited for creating an optical centrifuge. To explore this, an integrated EOSpace polarization controller with response times of nanoseconds was used~\cite{noauthor_polarization_nodate}. 

However, as the polarization state is a non-linear function of an applied voltage, a careful calibration is required to determine the input voltages that correspond to generating a desired polarization as a function of time. The device is also sensitive to temperature fluctuations and movements of the connecting fibers. As such, a protocol has to be devised to determine the right voltages to obtain the polarizations that are desired to achieve the acceleration.

\begin{figure}
    \centering
    \includegraphics[width=0.5\textwidth]{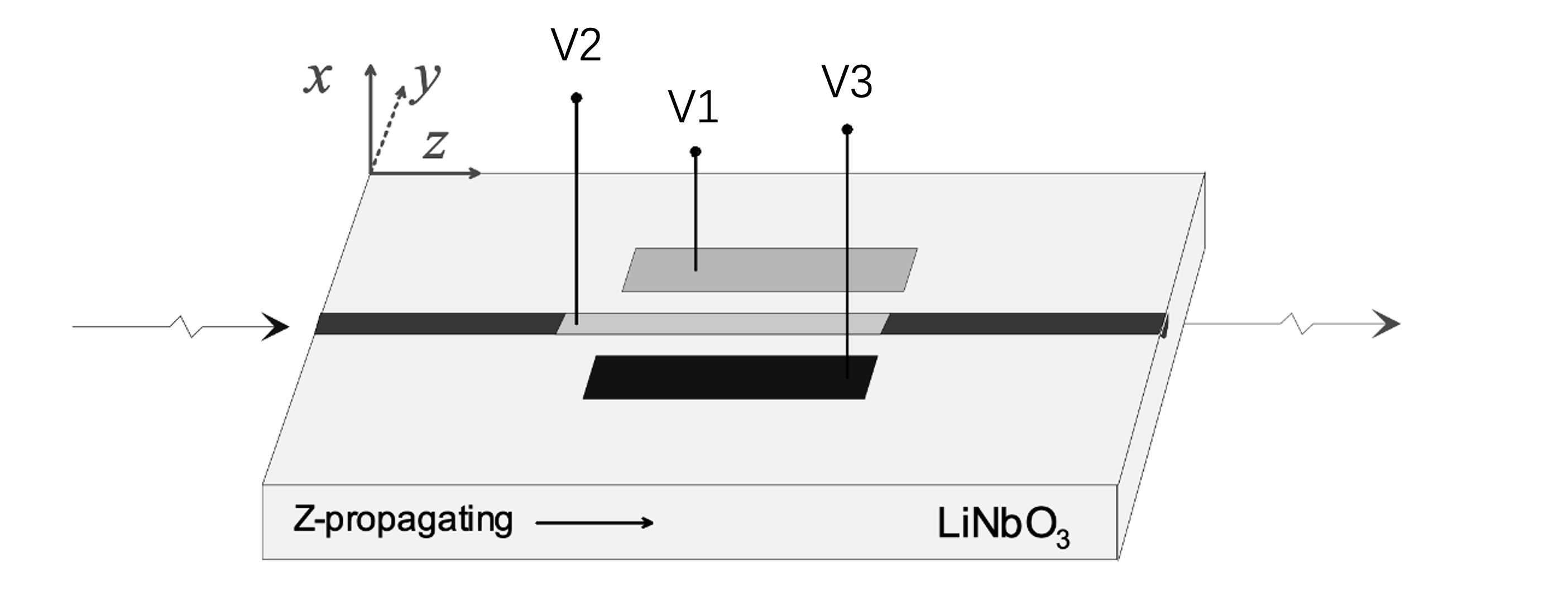}
    \caption{A schematic diagram of an electro-optic polarization modulator made from a birefringent LiNbO$_2$ waveguide. Also shown are the electrodes placed around the waveguide which allow the refractive index of this material to be rapidly controlled with the application of voltages to the three electrodes shown. This electrically tunable polarization controller can in principle be used to create any polarization state. In particular it is suitable for generating light of linear polarization with a monotonically increasing angle of polarisation. We adapted this figure from~\cite{eosgraph}.}
    \label{fig:EOSdetail}
\end{figure}

We characterize an EOSpace polarization controller by determining the polarization state of the output light as a function of the voltage applied to the electrodes labelled $V_1$ and $V_2$ in Fig. \ref{fig:EOSdetail}. The terminal $V_3$ is unused. We use a polarimeter to obtain a map of the Stokes parameters obtained by applying a pair of voltages across the polarization controller. A schematic of the polarization controller is shown in Fig. \ref{fig:EOSdetail}. 

\begin{figure}
    \centering
    \includegraphics[width=0.4\textwidth]{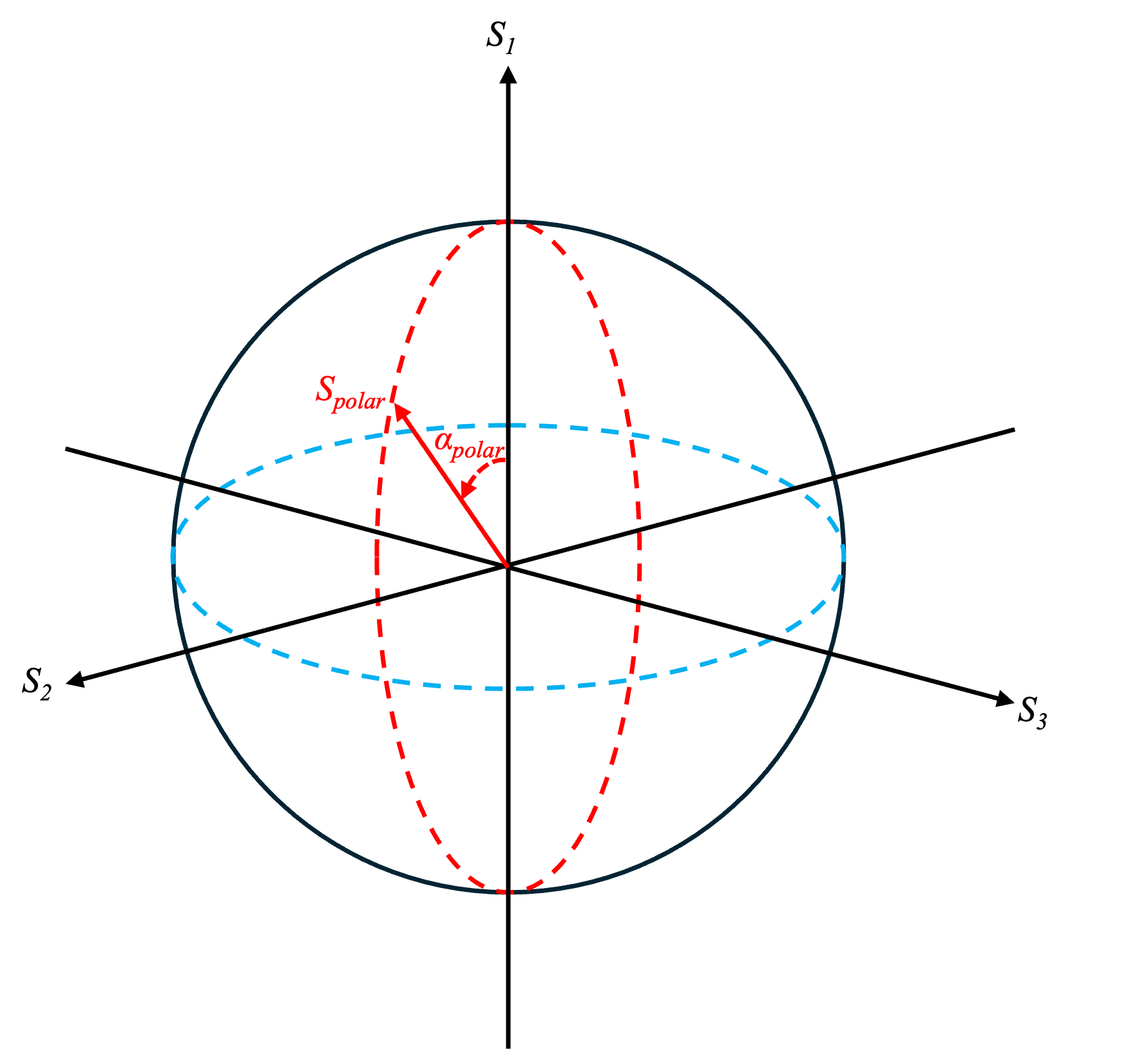}
    \caption{Stokes parameters ($S_1,S_2,S_3$) represented as a vector (red) on the Poincaré sphere. Linear polarization is defined by the vector lying within the plane defined by ($S_1,S_2$) with the angle of polarization given by $\alpha_{polar}$.}
    \label{fig:stocks_sphere}
\end{figure}

The Stokes parameters $S = (S_0, S_1, S_2, S_3)$ are shown in Fig. \ref{fig:stocks_sphere}, and can be written as a function of voltages $V_1, V_2$ applied, so $S=S(V_1, V_2)$. For a normalised Stokes parameter, $S_0=1$, we only require the three Stokes parameters, $S_1(V_1,V_2)$,  $S_2(V_1,V_2)$ and $S_3(V_1,V_2)$. The variation of Stokes parameters with applied voltage must be measured. As the output polarization varies with the voltages applied on $V_1$ and $V_2$, a mapping of this parametric surface can be performed via measuring the output polarization as a function of various input voltages \cite{hidayat2008fast_nonlinear}.

\begin{figure}
    \centering
    \includegraphics[width=0.5 \textwidth]{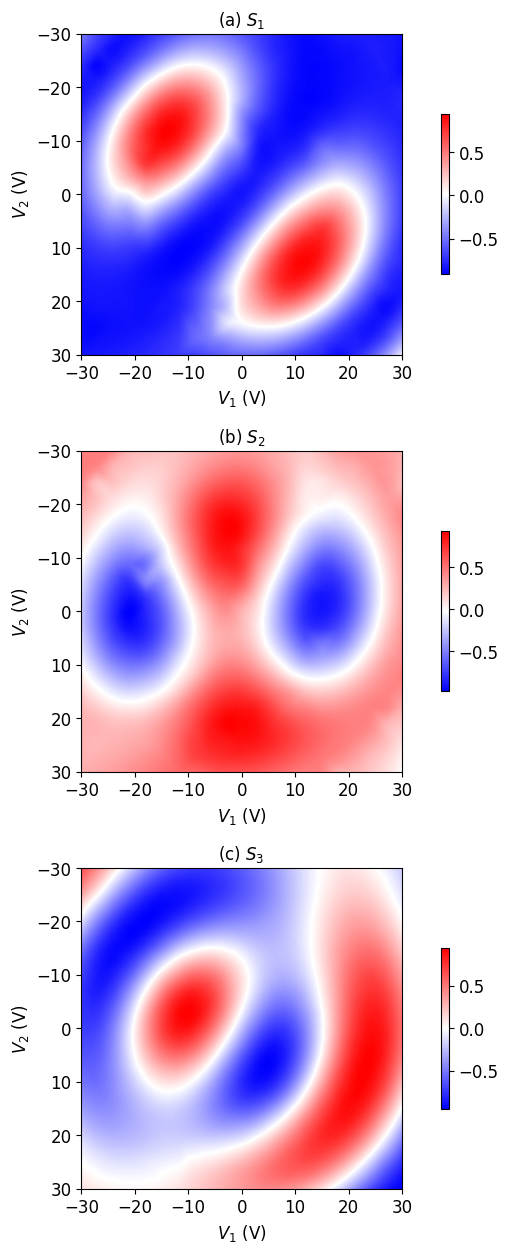}
    \caption{The interpolated voltage maps with $21\times21$ points. Each voltage generates a Stokes vector. Upon normalising the parameters by dividing each with $S_0$, there are 3 unique values where (a) is $S_1$, (b) is $S_2$ and (c) is $S_3$. One can picture each parameter as a function of $V_1$ and $V_2$, such as $S_1=S_1(V_1,V_2)$.}
    \label{fig:map21}
\end{figure}

To understand the variation in Stokes parameters with applied voltage we measure the Stokes parameters as a function of $V_1$ and $V_2$ over a range of $\pm$30 volts. Fig. \ref{fig:map21} are two dimensional maps of each Stokes parameter as a function of the applied voltages. Most of our measurements are performed using $21\times21$ equally spaced voltage values. We then use interpolation between these grid values to determine the Stokes parameters for any applied voltages.

\begin{figure} 
    \centering
    \includegraphics[width=0.5\textwidth]{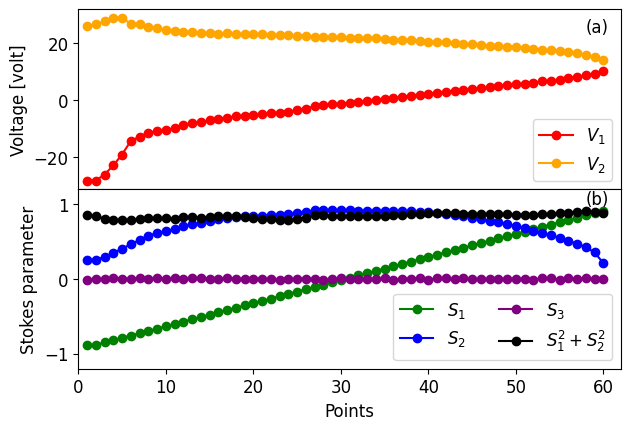}
    \caption{(a) The selected voltage pairs by using the constraints where $-0.01\leq S_3\leq 0.01$ and $0.78 \leq S_1^2 + S_2^2\leq 1$ to rotate linear polarization. (b) The corresponding Stokes parameters. }
    \label{fig:volt_s}
\end{figure}

Fig. \ref{fig:map21} shows the voltage maps of the Stokes parameters with interpolation. For angular acceleration, we ideally require a linearly polarized beam such that $S_3=0$, and $S_1^2 + S_2^2 = 1$, in order to find appropriate voltage pairs that can be used for the required chirped polarization. To do this, we first find the points where $S_3\approx0$ and $S_1^2 + S_2^2\approx1$ and also where $S_2 >0$. In order to find voltage pairs  where the voltage can monotonically vary,  we relax the constraints such that $-0.01\leq S_3\leq 0.01$ and $0.78 \leq S_1^2 + S_2^2\leq 1$. This produces a slightly elliptically polarized beam, which as we will show below, is sufficient for the required angular acceleration. We then sort the Stokes parameters in order of increasing polarization angle where $\alpha_{polar} = \cos^{-1}{(S_1/\sqrt{S_1^2 + S_2^2})}$. We order the points from minimum to maximum and search for the points that obey $0\leq S_1 \leq 1$ using the 60 points as shown in Fig. \ref{fig:volt_s}(a).  We check that the voltage pairs for each angle are changing monotonically and in small increments, the function generator, amplifier and modulator all have a finite slew rate. The variation in the Stokes parameters for these 60 voltage pairs is shown in Fig. \ref{fig:volt_s}(b). Note that the light is not completely linearly polarized and contains some ellipticity. We only need to rotate the polarization by 180 degrees since alignment of the molecule does not differentiate between linear polarization that is shifted by $\pi$ degrees. 

\begin{figure}
    \centering
    \includegraphics[width=\columnwidth]{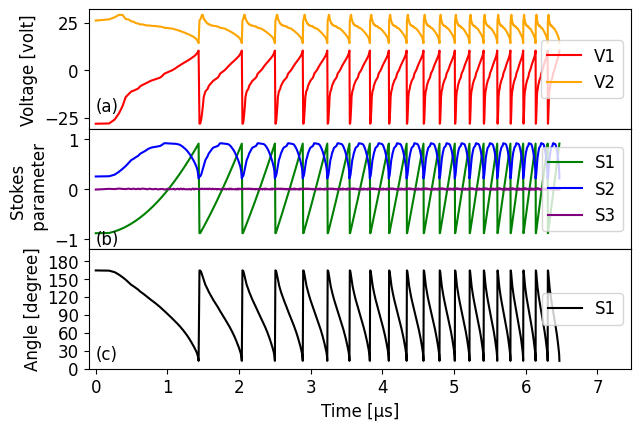}
    
    \caption{(a) The variation of voltage with chirp $10^{12}$ s$^{-2}$ and (b) the corresponding Stokes parameters and (c) the rotational angle of the linear polarization.}
    \label{fig:chirp_volt}
\end{figure}

The voltage pairs required to achieve a chirped polarization vector with a value of $10^{12}$ s$^{-2}$ are illustrated in Fig. \ref{fig:chirp_volt}(a). The chirp involves repeatedly applying the procedure depicted in Fig. \ref{fig:volt_s}, with progressively increasing rates of change. The corresponding evolution of the Stokes parameters throughout this accelerating sequence is presented in Fig. \ref{fig:chirp_volt}(b). Additionally, the rotation angle of polarization along the $x$ direction during this process is shown in Fig. \ref{fig:chirp_volt}(c). Our simulation results indicate that although the achieved rotation is not precisely over 180 degrees, the generated polarization is sufficient for the creation of a stable optical centrifuge.


\begin{figure} 
    \centering
    \includegraphics[width=0.48\textwidth]{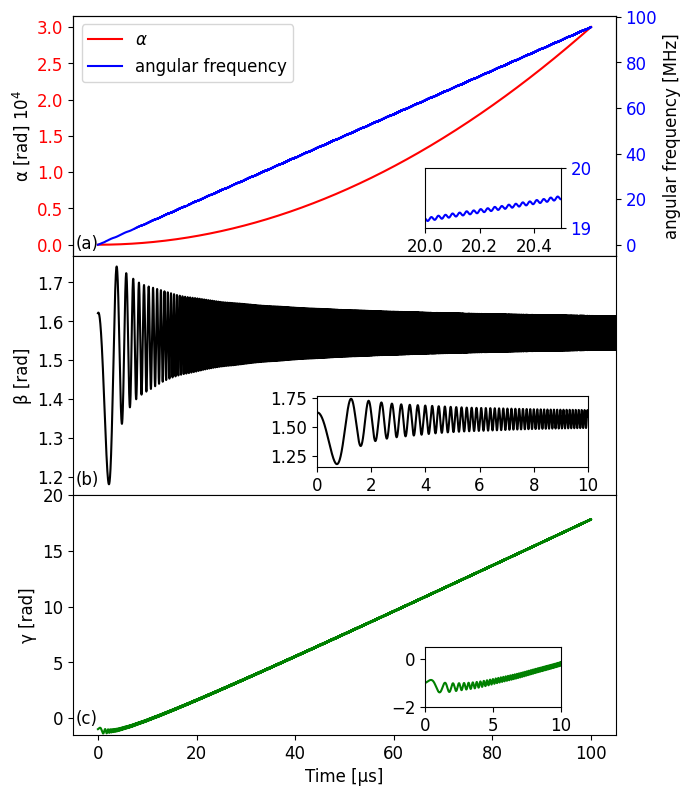}
    \caption{(a) The rotational angle $\alpha$ as a function of time as well as the angular frequency when the nanorotor is initially aligned with Stokes vector $S_1$ and is 0.05 rad shifted from equilibrium point in $\beta$ direction and the initial angle in $\gamma$ direction is -1 rad. We assume the nanorotor is initially cooled to 1 K in all degrees of freedom. (b) The rotation angle $\beta$ as a function of time. (c) The rotation angle $\gamma$ as a function of time.}
    \label{fig:stoke_alpha}
\end{figure}

The simulated operation of a centrifuge using the polarization control is shown in Fig. \ref{fig:stoke_alpha}, where we use the Stokes parameters as input to the equation \ref{dphidt} to simulate the angle of alignment $\alpha$ as a function of time. The nanoparticle's $\alpha$ axis is initially aligned with the Stokes vector $S_1$ so that $\alpha$=-0.365 and the $\beta$=1.6208 axis is initially at 0.05 rad shifted from its equilibrium point of $\beta_0 = \pi/2$ and the initial angle of $\gamma$ is -1 rad. We also assume the nanorotor is initially cooled to 1 K in all degrees of freedom~\cite{dania2021opticalCooling1,PhysRevResearch.6.033009Cooling2,pontin_simultaneous_2023,bang_five-dimensional_2020}. Our calculations indicate that the nanoparticle adiabatically follows the polarization vector $S_1$ and can be rotated up to 100 MHz using the same Stokes parameters as shown in Fig. \ref{fig:stoke_alpha}(a). Here we show both $\alpha$ and the angular frequency as a function of the acceleration duration of 100 $\upmu$s.  Fig. \ref{fig:stoke_alpha}(b) shows the evolution of the angle $\beta$ during acceleration. It shows the reduction in its amplitude due to the gyroscopic coupling between the $\alpha$, $\beta$ and $\gamma$ motion limits $\beta$ to $ \pi/2$ for long accelerations, the equilibrium angle for the torsional motion of this degree of freedom. Fig. \ref{fig:stoke_alpha}(c) is a similar plot for $\gamma$ which is drived by the polarization with the same rotational frequency, where $\gamma$ is largely unaffected by the chirping process.

\section{Applications}
An optical centrifuge could be used to explore the quantum nature of rotation.  For example, for a rigid body with three distinct principal moments of inertia, the rotation of the body around its intermediate axis is not stable and induces a phenomena called the tennis racket or Dzhanibekov effect.  In the quantum regime, this effect can be seen as a tunneling between orientational states. The quantum tennis-racket effect in nanorotors has been shown to exhibit higher coherence than for the classical motion and could be used to observe the quantum nature of rotation on mesoscopic-scale rotors~\cite{ma_quantum_2020}.  Here the centrifuge could be used to spin the particle to the desired frequency and the tennis racket effect could be observed over a short time periods with the optical potential turned off. Snapshots of the rotational state could be made by turning the tweezers back on, which would also allow the particle to be recaptured before repeating the process to measure the coherence time. An optical centrifuge could also be used to explore the Barnett effect on the nanoscale, a phenomenon in which spinning the particle can change the magnetization of a ferromagnetic particle \cite{stickler_quantum_2021,ma_quantum_2020}. 

As the separation between rotational states increases with rotational frequency, an optical centrifuge offers the possibility to resolve this quantization. Potentially, the quantum nature of nanoparticle rotation could be observed by measuring this discreteness when placed in very high rotational states. This should be observable as the rotational frequency spacing increases linearly with rotational quantum number. By using symmetric nanorotor with $J_\alpha= 6.6\times10^{-34}$ kg m$^2$/s the nanoparticle could be rapidly accelerated up to rotational states as high as J=15,000. This corresponds to rotational frequencies in the MHz range well below the GHz range that have been observed~\cite{ahn_optically_2018}. The separation between rotational states is of the order of 400 Hz which should be easily resolvable in an experiment. Using the optical centrifuge, the nanorotor could be driven and maintained at a well-defined rotational frequency. The ability to spin nanoparticles to well-defined frequencies whose value is not defined by dissipation could be used for levitated gyroscopes which have already shown promise~\cite{Zeng2024}.

\section{Conclusions}
We have studied the creation of an optical centrifuge for the controlled rotation of nonspherical nanoparticles (nanorotors) in a standard optical tweezers using a chirped acceleration of the tweezer polarization. We have shown that rotational acceleration of a nanorotor up to at least 100 MHz is feasible using this mechanism if the particles center-of-mass is cooled so that it remains close to the trap center. As part of this study, we have established the limits on the rotational acceleration set by the laser trap parameters, as well as the nanoparticle's properties and initial orientation with respect to the polarization vector. Subsequently, we simulated the optical centrifuge's process demonstrating its operation on all angular degrees of freedom ($\alpha,\beta,\gamma$). The effect of collisions was also simulated and high rotational speeds can still be achieved in modest vacuum. Realistic parameters based on the fast polarization rotation from an electro-optic modulator were explored. These experiments showed that even non-optimal polarization control should allow for a rapid acceleration of a nanorotor to the desired rotational rate of approximately 100 MHz without loss from the accelerated potential.

\clearpage

\section*{Acknowledgements}
PX, KKKH, MR, JMHG and PFB acknowledge funding from the EPSRC and STFC via Grant Nos. EP/N031105/1, EP/S000267/1, EP/W029626/1, EP/S021582/1 and ST/W006170/1. 

\bibliography{centrifuge}

\end{document}